\documentclass[12pt]{article}
\usepackage{amssymb}
\usepackage{epsfig}
\usepackage{psfrag}

\newcommand{\beqn}{\begin{eqnarray}}
\newcommand{\beq}{\begin{equation}}
\newcommand{\eeqn}{\end{eqnarray}}
\newcommand{\eeq}{\end{equation}}

\newcommand{\tr}{\mathop{\rm Tr}}

\newcommand{\abstracts}[1]{{
\centering{\begin{minipage}{13.0truecm}
\normalsize\baselineskip=15pt \centerline{\footnotesize
ABSTRACT}\vspace*{0.3cm}
\parindent=20pt #1
\end{minipage}}\par}}

\begin{document}

\hfill HU-EP-05/33, ITEP-LAT/2005-12~~~~~~~~ 

\hfill July 2005~~~~~~~~

\begin{center}
{\baselineskip=24pt {\Large \bf Universality check of Abelian Monopoles} \\

\vspace{1cm}

{\large
V.~G.~Bornyakov$^{\rm a,b}$, E.--M.~Ilgenfritz $^{\rm c}$ 
and M.~M\"uller--Preussker $^{\rm c}$
}
\vspace{.5cm}
{\baselineskip=16pt
{ \it

$^{\rm a}$ Institute for High Energy Physics, Protvino 142284,
    Russia\\
$^{\rm b}$ Institute of Theoretical and  Experimental Physics,
    Moscow, 117259, Russia\\
$^{\rm c}$ Humboldt--Universit\"at zu Berlin, Institut f\"ur Physik, 
    D--12489 Berlin, Germany
}
}}
\end{center}
\abstracts{
We study the Abelian projected $SU(2)$ lattice gauge theory after gauge 
fixing to the maximally Abelian gauge (MAG).
In order to check the universality of the Abelian dominance 
we employ the tadpole improved tree level (TI) action. We show that the 
density of monopoles in the largest cluster (the IR component) is finite 
in the continuum limit which is approximated 
already at relatively large lattice spacing. 
The value itself is smaller than in the case of Wilson action.
We present results for the ratio of the Abelian to non-Abelian string tension
for both Wilson and TI actions for a number of lattice spacings in the range
$0.06~\mbox{fm} < a < 0.35~\mbox{fm}$. These results show that the ratio   
is between 0.9 and 0.95 for all considered values of lattice couplings and
both actions. 
We compare the properties of the monopole clusters in two gauges - in MAG and
in the Laplacian Abelian gauge (LAG).
Whereas in MAG the infrared component of the monopole 
density shows a good convergence to the continuum limit, we find that in LAG 
it is even not clear whether a finite limit exists.
 }
\section{Introduction}

The dual superconductor scenario of confinement has received support
from many observations made as well in gluodynamics \cite{review1,review2} 
as in full lattice QCD \cite{Bornyakov:2003vx}.
The most intensively investigated case was $SU(2)$ gluodynamics. The scaling
properties of many gauge dependent observables such as the Abelian string 
tension, the effective monopole action, the monopole density etc. have been checked.
It has been shown that some properties of the monopoles in MAG can be 
explained by percolation theory or by free particle field theory 
\cite{Chernodub:2002tw}. Despite this progress there is lack of universality 
checks, {\it i.e.} the independence of the choice of action has not been confirmed. 
Apart from papers  \cite{Poulis:1997zx} and \cite{Suzuki:2004uz}, 
always the Wilson action for the gauge field has been employed. 
In comparison with Ref. \cite{Poulis:1997zx}, where the same TI action 
was used, we have much better statistics and better gauge fixing, i.e.
lower effects of gauge fixing ambiguities. 
In \cite{Suzuki:2004uz}, where a different improved action was considered,
the study was made for one value of lattice spacing only and thus no scaling
studies were attempted. 
In the present paper we are aiming to make a contribution to closing this gap. 
The other, perhaps even more important problem is the dependence on the gauge 
condition used for Abelian projection. There are various opinions on this problem. 
Some authors believe that the occurrence of monopole condensation itself and, 
correspondingly, the dual superconductor properties of the vacuum have to exist 
and do exist in any Abelian gauge~\cite{Carmona:2001ja}.
On the other hand recent results \cite{Belavin:2004ss}
obtained with the Fr\"ohlich-Marchetti monopole creation operator show that
the monopole condensate depends on the choice of the Abelian gauge.
In \cite{Chernodub:2003mm} it was argued that in the Abelian gauge
defined by diagonalization of the Polyakov loop operator the condensation
of monopoles does not necessarily lead to formation of the Abelian
flux tube between static quarks.
We share the opinion, that any Abelian projection is made with the goal of 
separating degrees of freedom responsible for infrared physics, which thus 
should carry all low momenta of the original non-Abelian gauge field, from
ultraviolet degrees of freedom which are responsible for short distance
physics (if they have a sensible continuum limit) or might even be mere 
lattice artefacts. Such separation does not need to be accomplished in any 
conceivable Abelian gauge. Rather we expect that there might exist a class 
of gauges which indeed have this property.
The maximally Abelian gauge MAG is a very likely candidate to belong to this 
class, and the Laplacian Abelian gauge (LAG)~\cite{vanderSijs:1997hi} is widely 
considered as another good candidate. Since the analytical study of these 
gauges in the nonperturbative regime  is very difficult and has not been 
accomplished so far despite many recent attempts (for MAG studies 
see, e.g. \cite{Dudal:2004rx} and references therein), the numerical study is 
the only way to approach this problem in practice. Therefore, in this study we 
also compare some of the properties of these two gauges. From the above point of
view, the issue of universality, it turns out that MAG is really unique to 
allow the separation of scales attempted by Abelian projection.

The paper is organized as follows.
In section \ref{sect:simulation_details} we specify the technical tools, in
particular the improved action used here in contrast to the Wilson action
and the method of gauge fixing.
Then, in section \ref{sect:NA_string_tension}, we briefly report on the 
evaluation of the string tension for the purpose of calibrating the lattice
scale corresponding to the improved action.
Section \ref{sect:monopole_density_MAG} contains our observations concerning the
scaling properties of the monopole densities and their IR and UV components
for both actions in the maximally Abelian gauges.
In section \ref{sect:monopole_density_LAG} we show that similarly to MAG
the monopole clusters obtained in the Laplacian Abelian gauge might be splitted 
into IR and UV components but their scaling
properties are quite different from those observed in MAG. 
Section \ref{abstr} is devoted to the Abelian dominance
study. Our results indicate universality of the Abelian dominance in the 
continuum limit. Finally we summarize our findings in section \ref{summ}.

\section{Simulation details}
\label{sect:simulation_details}

To address again the question of universality, we employ here the tree level 
improved action of the form \cite{Alford:1995hw}
\beq 
    S =   \beta_{imp} \sum_{pl} S_{pl}
       - {\beta_{imp} \over 20 u_0^2} \sum_{rt} S_{rt}
\label{eq:improved_action}    
\eeq
where $S_{pl}$ and $S_{rt}$ denote plaquette and $1 \times 2$ rectangular
loop terms in the action,
\beq 
S_{pl,rt}\ = \ {1\over 2}{\rm Tr}(1-U_{pl,rt}) \, ,
\label{eq:terms}
\eeq
the parameter $u_0$ is the {\it input} tadpole improvement 
factor taken here equal to the fourth root of the average plaquette 
$P=\langle \frac{1}{2} {\mathrm tr} U_{pl} \rangle$.

In our simulations we have not included one--loop corrections to the coefficients, 
for the sake of simplicity and also to be able to compare with the results 
of Ref.~\cite{Poulis:1997zx} after making a few improvements in comparison with 
this work in other directions.
First, we improved substantially the gauge fixing as will be explained later. 
Second, we have now better statistics and worked on larger physical volumes.
This has allowed to determine more reliably various Abelian observables and their
infrared part.
Third, we have used a new smearing techniques which enabled us to make more precise 
measurements of the non-Abelian string tension. This was necessary to assess the
nonperturbative scaling of various monopole densities. 

We also make a comparison of these Abelian observables obtained in MAG 
and LAG, respectively. The MAG is fixed by the maximization of the lattice 
functional
\beq
 F(U) = \frac{1}{8 V} \sum_{n,\mu}
        \tr\left( \sigma_3 U_{n,\mu} \sigma_3 U_{n,\mu}^{\dagger} \right) \quad,  
\label{eq:max_func}
\eeq
\noindent
with respect to local gauge transformations
\beq
U_{n,\mu} \rightarrow U^{g} _{n,\mu} = g_n U_{n,\mu} g_{n+\mu}^{\dagger} \quad.    
\label{eq:gauge_trafo}
\eeq
For MAG we applied the simulated annealing algorithm. The details of the gauge 
fixing procedure can be found in \cite{Bali:1996dm}.
We have applied the algorithm to 10 randomly replicated gauge copies of each 
Monte Carlo configuration in the hope to find among the 10 local maxima one 
closest to the global maximum.
This procedure proved to be the best so far to fix MAG as well as to fix center 
gauges \cite{Bornyakov:2000ig}.
Although there is no proof we hope that our results for gauge noninvariant
observables are numerically close to those we would obtain evaluating it at
the gauge equivalent representant carrying the global maximum of 
(\ref{eq:max_func}) for every configuration of the gauge field.

Fixing the LAG amounts to finding the eigenvector with the lowest eigenvalue 
of the covariant Laplacian operator in the adjoint representation,
\beq 
 -\square_{\,nm}^{\,ab}=\sum_\mu\left( 2\delta_{nm}\delta^{ab}
  - R_{n,\mu}^{\,ab}\delta_{m,n+\hat{\mu}}
  - R_{n-\hat{\mu},\mu}^{ba}\delta_{m,n-\hat{\mu}}\right)
\label{eq:Laplacian}
\eeq
with the adjoint link variable
\beq 
R_{n,\mu}^{ab} = \tr\left( \sigma_a U_{n,\mu} \sigma_b
      U_{n,\mu}^{\dagger} \right) \, .
\label{eq:adjoint_link}
\eeq
The gauge transformation $g_n$ is then determined by the requirement to rotate 
this eigenvector $\phi^a_n$ to the 3$^{\mathrm rd}$ color axis at every 
site $n$:
\beq
\rho_n \sigma_3 = \sum_{a=1}^3\phi^a_n\ g_n \sigma_a g^\dagger_n, \,\,\, 
\rho_n = \sqrt{ \vec{\phi}_n^2} \, .
\label{eq:unitary_gauge}
\eeq

The simulations with the action (\ref{eq:improved_action}) 
have been performed with parameters 
given in Table~\ref{t1}.  The parameter $u_0$  
has been iterated over a series of Monte Carlo runs in order to match the 
fourth root of the average plaquette $P$.
The corresponding entries give an impression of the accuracy of matching.
\begin{table}[htb]
\begin{center}
\caption{Details of the simulations with improved action}\label{t1} 
\vspace{.3cm}
\setlength{\tabcolsep}{0.55pc}
\begin{tabular}{cccccc}
$\beta_{imp}$ & L & $N_{conf}$ & $u_0$ & $<P>^{1/4} $ & $\sqrt{\sigma a^2}$ \\ 
\hline
2.7 & 12 &  60 & 0.87164 & 0.87165(2)  & 0.60(5)   \\
3.0 & 12 & 200 & 0.89485 & 0.89478(2)  & 0.366(8) \\
3.1 & 12 & 200 & 0.90069 & 0.90069(1)  & 0.309(6)  \\
3.2 & 16 & 200 & 0.90578 & 0.905765(3) & 0.258(5)  \\
3.3 & 16 & 100 & 0.91015 & 0.910152(4) & 0.219(3)  \\
3.3 & 20 &  50 & 0.91015 & 0.910153(3) & 0.215(3)  \\
3.4 & 20 & 100 & 0.91402 & 0.914020(2) & 0.180(3)  \\
3.5 & 20 & 100 & 0.91747 & 0.917481(1) & 0.151(3)  \\
3.5 & 24 &  50 & 0.917475 & 0.917484(2) &0.152(2)  \\
\hline
\end{tabular}
\end{center}
\end{table}
For two values of $\beta_{imp}$ we simulated lattices of two sizes.
The value of the parameter $u_0$ fixed on smaller lattice was used as an input
for larger lattice simulations. The string tensions obtained on these
lattices are in agreement within error bars. The smaller (larger) lattices 
were used to study LAG (MAG).
\begin{table}[htb]
\begin{center}
\caption{Details of the simulations with Wilson action}\label{t2} 
\vspace{.3cm}
\setlength{\tabcolsep}{0.55pc}
\begin{tabular}{cccc}
$\beta$ & L & $N_{conf}$ & $\sqrt{\sigma a^2}$ \\ 
\hline
2.40& 32 &  35 &0.264(7) \\
2.45& 24 & 100 &0.226(3) \\
2.50& 24 & 100 &0.185(2)    \\
2.55& 28 & 100 &0.159(2) \\
2.60& 28 & 100 &0.1319(15)    \\
2.65& 32 & 40 &0.114(2)    \\
\hline
\end{tabular}
\end{center}
\end{table}

\section{The non-Abelian string tension}
\label{sect:NA_string_tension}

\begin{figure}[htb]
\begin{center}
\hspace{-2cm}\epsfxsize=13truecm \epsfysize=11truecm \epsfbox{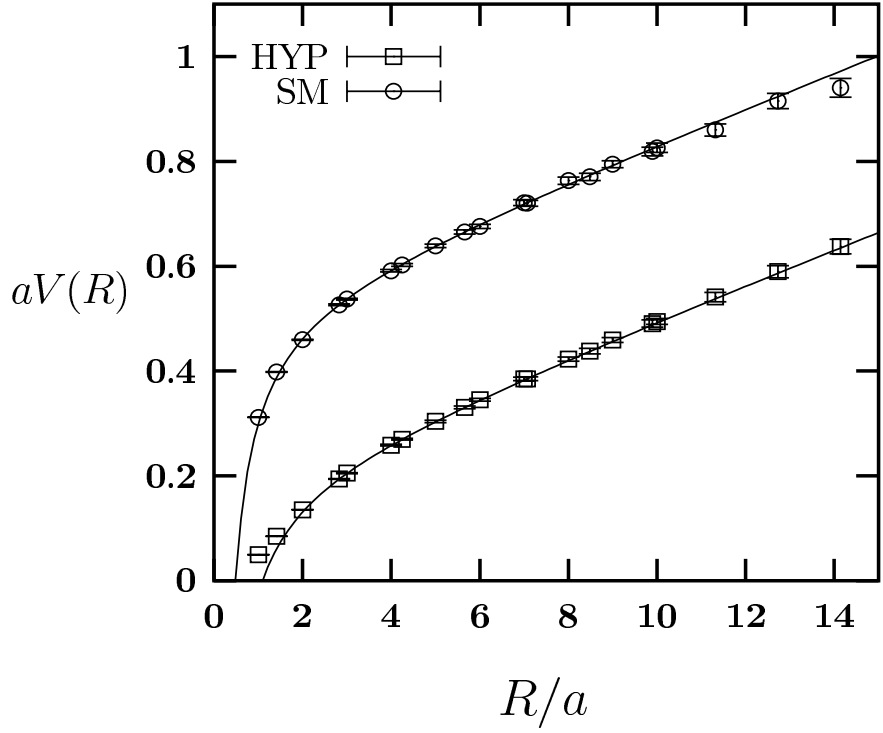}
\end{center}
\vspace{-1.5cm}
 \caption{Non-Abelian potentials for the TI action obtained without (circles, SM) 
and after (squares, HYP) hypercubic blocking (both with spatial smearing) 
vs. $R/a$ for $T/a=5$ at $\beta_{imp}=3.4$.}
\label{fig:hcb_poten2}
\end{figure}

In order to fix the physical lattice scale we need to compute one physical 
dimensionful observable the value of which is known. For this purpose we
choose the string tension $\sigma$. 
The string tension for action (\ref{eq:improved_action}) was computed 
long ago in \cite{Poulis:1997zx} but we will improve this measurement
according to present standards. 
We use the hypercubic blocking (HYP) invented by the authors
of Ref.~\cite{Hasenfratz:2001hp} to reduce the statistical errors. 
This method has been successfully applied to static potential calculations
in $SU(3)$ gluodynamics~\cite{Hasenfratz:2001hp, Hasenfratz:2001tw, 
Gattringer:2001jf} and in full lattice QCD at finite 
temperature~\cite{Bornyakov:2004ii}. After one step of HYP, about 20 sweeps of 
APE smearing \cite{APE} were applied to the space like links. The spatial 
smearing is made, as usually, in order to 
variationally improve the overlap with a mesonic flux tube state. 
In Fig.~\ref{fig:hcb_poten2} we compare potentials obtained with and without
HYP procedure. As was observed in the cited above papers the HYP potential 
differs essentially by a constant shift corresponding to reducing the
static source self-energy. One can see from the figure that HYP decreases both 
statistical errors and effects of rotational invariance breaking.
Since HYP changes the potential at 
small distances we included only distances $R/a > 2$ into our fits of 
the static potential. The resulting values for $\sqrt{\sigma a^2}$ are 
also included in Table~\ref{t1}.
The string tension was also calculated with Wilson action. In this case 
APE smearing for space links and the additional trick of link integration 
\cite{Parisi} for time links  were used in the evaluation of Wilson loops. 
The results for $\sqrt{\sigma a^2}$ and details of 
simulations with Wilson action are presented in Table~\ref{t2}.

\section{The monopole density}
\label{sect:monopole_density}
\subsection{MAG}
\label{sect:monopole_density_MAG}
 
After fixing the Abelian gauge the Abelian projection can be made:
\beq
U_{n,\mu} = C_{n,\mu} u_{n,\mu}\quad ,
\label{coset}
\eeq
where the Abelian field is contained in
$u_{n,\mu} = \mbox{diag}( e^{i \theta_{n,\mu}},e^{-i \theta_{n,\mu}})$, 
$\theta_{n,\mu} \in (-\pi,\pi]$, and $C_{n,\mu}$ is the coset field
describing charged gluons. 
The Abelian plaquette angle 
\beq
\theta_{n,\mu\nu} = 
\partial_{\mu} \theta_{n,\nu}-\partial_{\nu} \theta_{n,\mu}
\eeq
is decomposed  into regular and singular parts:
\beq
\theta_{n,\mu\nu} = \overline{\theta}_{n,\mu\nu} + 2\pi m_{n,\mu\nu}\,,
\quad\quad \overline{\theta}_{n,\mu\nu} \in (-\pi,\pi]
\label{split}
\eeq
$\overline{\theta}_{n,\mu\nu}$ is a physical Abelian flux through the 
lattice plaquette $\{n,\mu\nu\}$, and $m_{n,\mu\nu}$ counts the number of Dirac 
strings through this plaquette.
The magnetic currents are then defined as follows:
\beq
k_{n,\mu}= 
\frac{1}{2}\epsilon_{\mu\nu\alpha\beta}\partial_\nu \overline{\theta}_{n,\alpha\beta} 
= -  \frac{1}{2}\epsilon_{\mu\nu\alpha\beta}\partial_\nu m_{n,\alpha\beta}
\eeq

We will measure the monopole density in lattice units $\rho^{lat}$  
defined as
\beq
\rho^{lat} = \sum_{n,\mu} \frac{|k_{n,\mu}|}{4 L^4}
\eeq
Since monopoles are three dimensional objects their physical density is 
related to the lattice density by $a^3 \rho = \rho^{lat} $.
With Wilson action, the first measurement of the monopole density in MAG 
gauge has been made in \cite{Bornyakov:1990es} with participation of the 
authors of the present paper.
That result was interpreted in the sense of asymptotic scaling.
In fact, the observation of asymptotic scaling at $\beta$ values in the 
range from 2.4 to 2.6 would seem rather strange today. 
It is known that the string tension does not follow the two--loop 
renormalization group formula over this range in $\beta$, such that 
the result obtained in \cite{Bornyakov:1990es} actually
implies the divergence of the monopole density with $\beta \rightarrow \infty$.
The situation has been partially clarified by Hart and 
Teper~\cite{Hart:1997vb,Hart:1999gr}. 
These authors found that on large enough lattices the network of magnetic currents
in each configuration consists of one large cluster and many other clusters 
with much smaller size.
The spectrum of cluster sizes falls into two very distinctive parts, 
disconnected by a gap. The density of currents forming the largest 
(percolating) cluster $\rho_{IR}$ has been measured 
in units of the string tension and a first indication of scaling of the ratio
$\rho_{IR}/\sigma^{3/2}$ 
has been found. More accurate measurements~\cite{Bornyakov:2001ux} have 
corroborated this kind of scaling behavior. The continuum limit for this ratio 
was determined as $\rho_{IR}/\sigma^{3/2}=0.65(2)$. 

Another important result 
obtained in Refs.~\cite{Hart:1997vb,Hart:1999gr} was the observation that the 
largest cluster alone produces almost the full monopole string tension. 
This fact has allowed the authors of Ref.~\cite{Bornyakov:2001hs} 
to call the monopoles belonging to this cluster ``infrared monopoles'' (IR) 
while the monopoles from the remaining clusters were called ultraviolet (UV) 
monopoles, implying that these monopoles are not relevant for IR physics. We 
should mention that, despite the fact that they are not relevant for the 
{\it confining} properties 
of the vacuum as supported by numerical observation, their relevance for the 
topology and therefore for the chiral properties of the vacuum has not yet been 
explored. We will keep (and have already used) the above notation, 
quoting IR and UV monopoles in the
following. In Ref.~\cite{Bornyakov:2001ux} it has been demonstrated that the 
density of UV monopoles, hence the total density, diverges in the continuum 
limit.  

It should be noticed that in \cite{Hart:1997vb,Hart:1999gr} a single,
supposedly percolating,
cluster with a size much larger than all other clusters in the given 
configuration was only found on large enough lattices. For decreasing lattice 
size $L$ and fixed lattice spacing this gap disappears, {\it i.e.} the 
largest and the second-largest cluster are of similar size. 
Consequently, the important property that the largest cluster alone produces 
almost the full monopole-related string tension, is lost.
This implies ``splitting'' of the largest cluster when the lattice 
volume decreases. It was also found in \cite{Hart:1997vb,Hart:1999gr} that the 
critical value of the lattice size $L_{crit}$, below which the largest cluster 
splits, is in the range of lattice sizes, where physical quantities do not 
show large finite volume effects, and, moreover, $L_{crit}$, measured in 
physical units, increases with decreasing lattice spacing. Such behavior
implies that in the continuum limit the gap in the spectrum of cluster sizes
might disappear and a clear separation of IR and UV monopoles may become 
impossible.

The solution of this problem was suggested in \cite{Bornyakov:2001ux}.
It was found empirically that the splitting of the largest cluster leads to
formation of clusters (rare for large enough lattices) with nonzero winding
\beq
w_\mu = \frac{1}{L_\mu} \sum_{k_{x,\mu} \in cluster} k_{x,\mu}\,.
\eeq
Such clusters might be very large on given configurations or be of moderate 
size. In both cases they extend through
the whole lattice at least in one direction and thus should be considered as 
relevant for the infrared physics. Furthermore, two or more of such clusters 
(forming together a combination of clusters with zero total winding $w_\mu$) 
might form boundaries of the same Dirac sheet, which is closed in one or more 
directions due to periodic boundary conditions. Let us note that in case of 
two wrapping clusters present in one configuration they unambiguously form
the boundary of the same Dirac sheet. When three wrapping clusters are present they
also form the boundary of one Dirac sheet or boundaries of two Dirac sheets.
In the latter case one of the clusters forms part of the boundaries of 
both Dirac sheets while two others form part of the boundary of one of those
Dirac sheets. 
It is natural to consider such clusters as one cluster when it concerns
the determination of clusters relevant for IR physics. It is also clear that
the splitting phenomena can be, at least partially, ascribed to the 
annihilation of parts of the percolating cluster 
through boundary conditions leading to formation of two disconnected clusters 
which still form a boundary of one Dirac sheet. 

Based on these observations, it was suggested in \cite{Bornyakov:2001ux} to 
define, for each configuration, a single IR cluster 
as the union of
the largest cluster (which might have trivial or nontrivial winding) and 
all clusters with nonzero winding. Numerical evidence was further presented 
in \cite{Bornyakov:2001ux} that under such definition
the size of the largest cluster changes smoothly with
the lattice size for physically large lattices.
There are preliminary results of ours to be published elsewhere, showing 
that with this definition the IR cluster alone reproduces almost the
full monopole-related string tension. We will use this definition in what 
follows.

As for lattice fields generated with TI action (\ref{eq:improved_action}), 
the monopole density has been measured by Poulis~\cite{Poulis:1997zx}. 
He concluded that the total density has correct scaling in the continuum limit. 
In the light of the discussion above this would mean that the TI action would
take an exclusive role. However, as we show below, this conclusion was wrong. 
\begin{figure}[hpbt]
\includegraphics[width=7.9cm,height=7cm,angle=0]{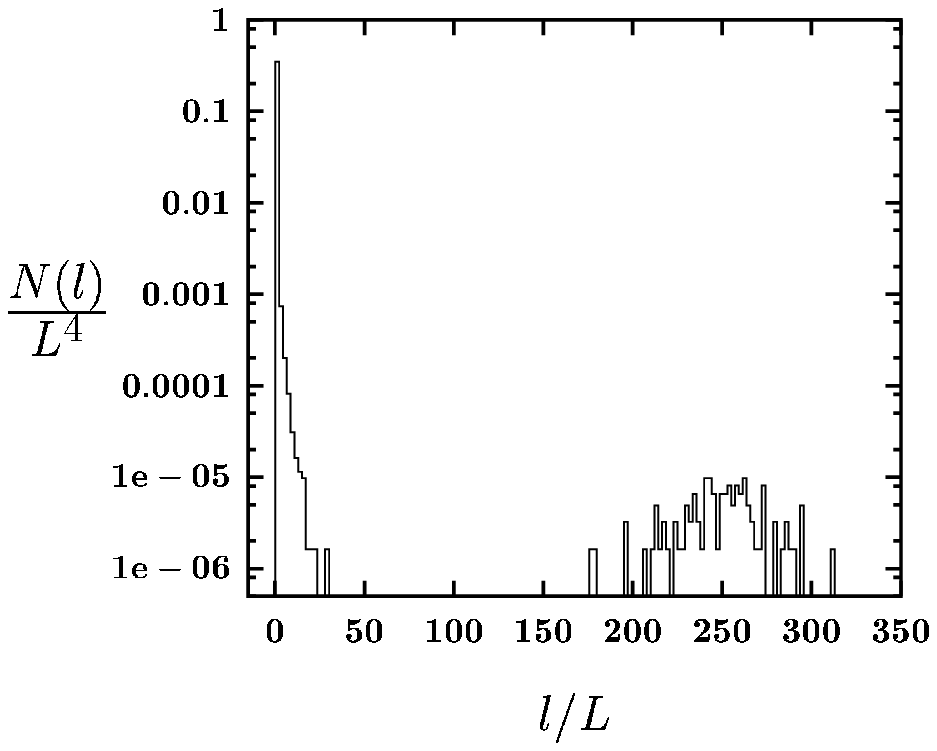}   
\includegraphics[width=7.9cm,height=7cm,angle=0]{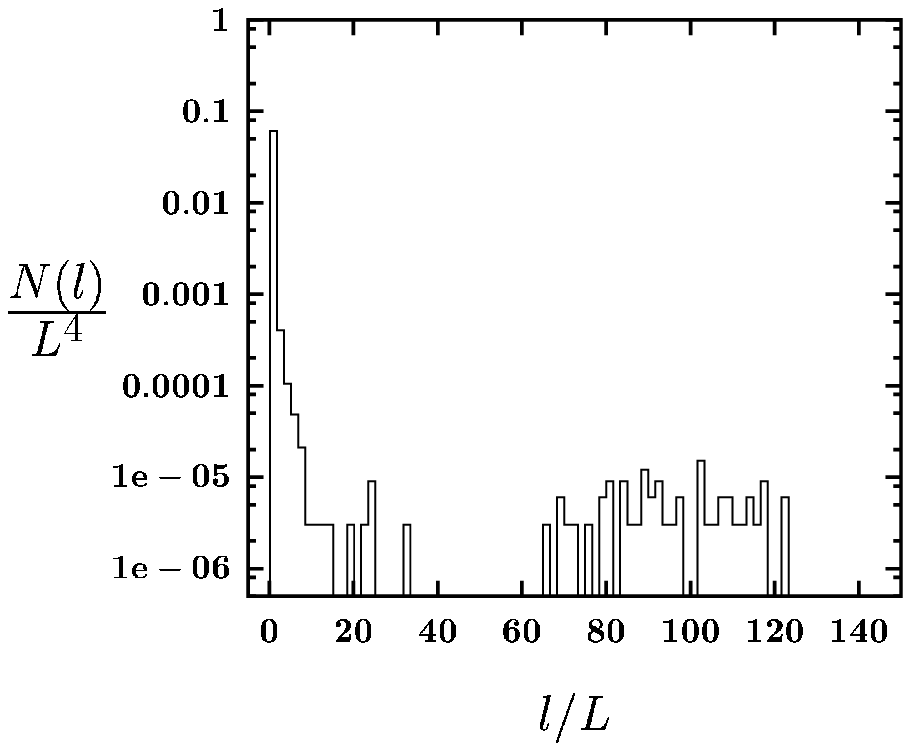}
\caption{The monopole cluster length distribution $N(l)$ 
at $\beta=2.55$ for the Wilson action (left) and at $\beta_{imp}=3.5$ for the 
TI action (right). }
\label{fig:dist255}
\end{figure}

In Fig.~\ref{fig:dist255} we show the distributions of the monopole clusters
length for Wilson action at $\beta=2.55$ and for TI action at $\beta_{imp}=3.5$.
Note, that 
a single IR cluster per configuration has entered the histograms which has been
defined for the configuration at hand as described above.
One can see that these distributions are qualitatively similar, i.e. for the TI 
action we also observe a clear splitting of the clusters into IR and UV clusters.  
\begin{figure}[hptb]
\begin{center}
\includegraphics[width=12.0cm,angle=0]{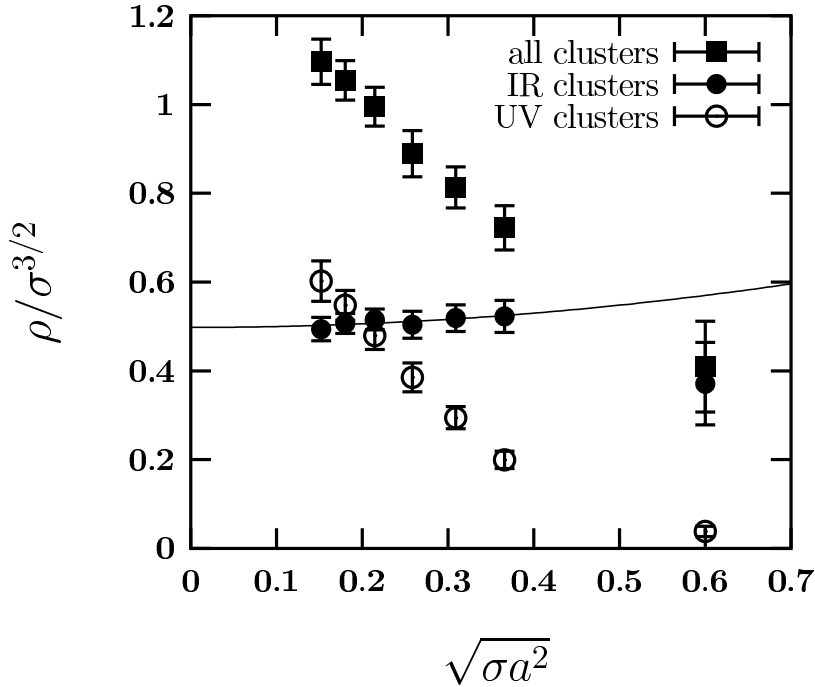}
\end{center}
\vspace{-1cm}
 \caption{The monopole densities in MAG for the case of the TI action. 
 The dashed line shows a quadratic fit.}
\label{fig:mondens1}
\end{figure}

Our results for the densities of IR and UV monopoles and the total density,
taken in lattice units, are presented in Table~\ref{t2}.
In physical units of $\sigma^{3/2}$, the results are shown as a function
of lattice spacing (in units of $\sqrt{\sigma a^2}$) 
in Fig~\ref{fig:mondens1}. One can see that the IR density converges to 
\begin{table}[htbp]
\begin{center}
\caption{Density of monopoles}\label{t3} 
\vspace{.3cm}
\setlength{\tabcolsep}{0.55pc}
\begin{tabular}{cccccccccc}
$\beta_{imp}$ & $\rho^{MA}_{tot}$ & $ \rho^{MA}_{IR}$ & $ \rho^{LA}_{tot}$ & $ \rho^{LA}_{IR}$ \\
\hline
2.7  & 0.08845(25) & 0.08014(30) & 0.1103(5) & 0.1032(5) \\
3.0  & 0.03537(16) & 0.02560(23) & & \\
3.1  & 0.02397(15) & 0.01529(20) & 0.0387(3) & 0.0282(3) \\
3.2  & 0.01534(11) & 0.00869(14) & & \\
3.3  & 0.00982(6)  & 0.00509(8)  & 0.0200(2) & 0.0118(2) \\
3.4  & 0.00618(4)  & 0.00297(5)  & & \\
3.5  & 0.00384(3)  & 0.00170(4)  & 0.0103(2) & 0.0049(2) \\
\hline
\end{tabular}
\end{center}
\end{table}
a finite value in the limit $a \rightarrow 0$. In contrast to this, the density 
of UV monopoles, and thus the total density, behaves divergent in the
continuum limit. We thus find that the TI action leads qualitatively to the 
same picture as was observed before with the Wilson action. 

\begin{figure}[htbp]
\begin{center}
\hspace{-1.cm}\includegraphics[width=8.5cm,angle=0]{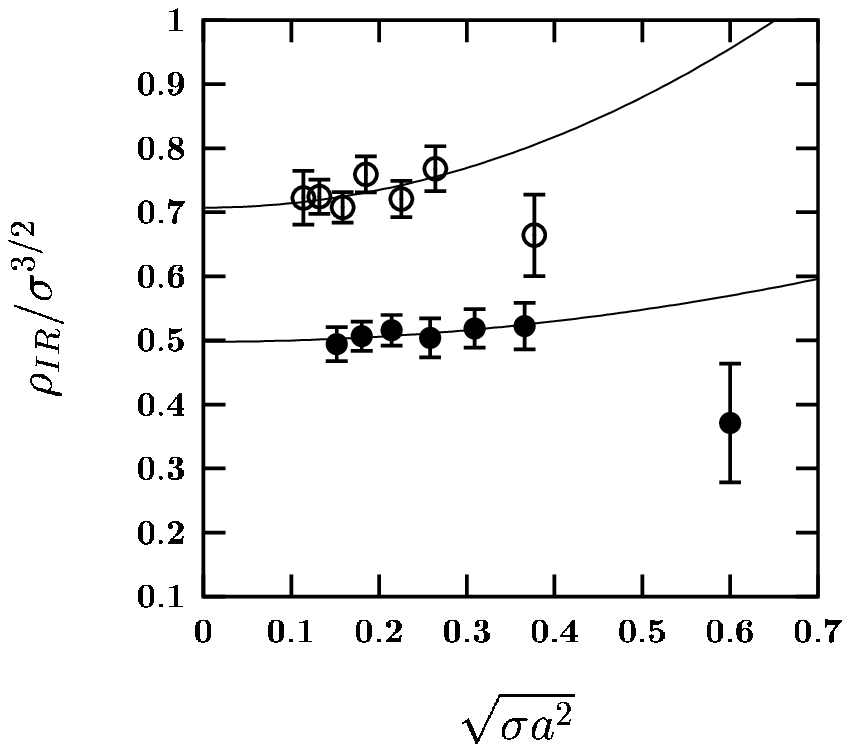}
\hspace{-.5cm}\includegraphics[width=8.5cm,angle=0]{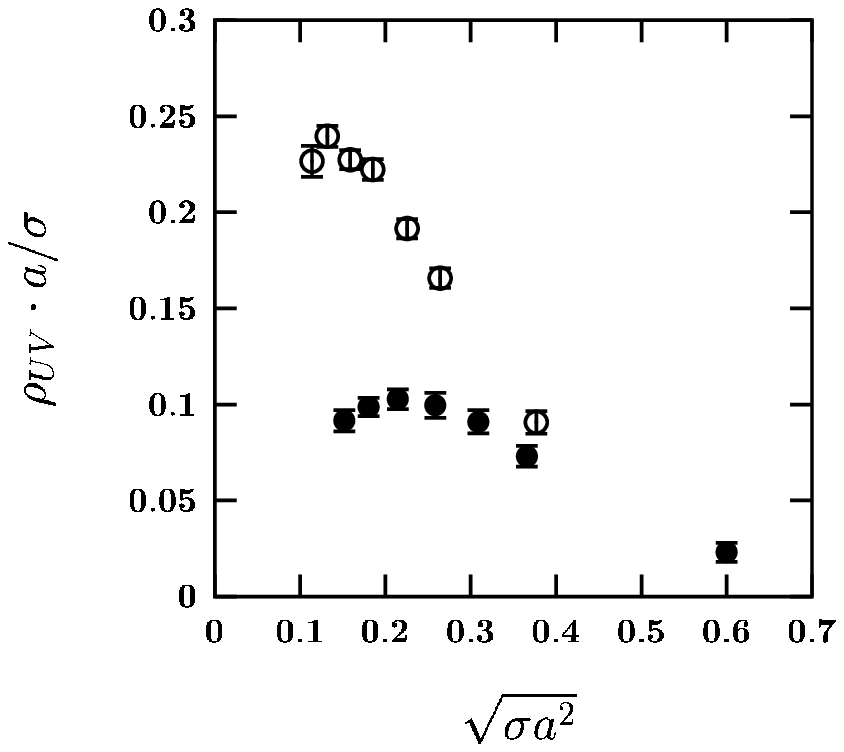}
\end{center}
\vspace{-1cm}
 \caption{Comparison of the monopole densities obtained with TI (full symbols)
 and Wilson actions (open symbols) in MAG: left - IR monopole density; 
 right - UV monopole density.}
\label{fig:mondens2}
\end{figure}

To make a quantitative comparison we plotted in Fig.~\ref{fig:mondens2} 
$\rho_{IR}$ (left) and  $\rho_{UV}$ (right) for two actions.
From these figures one can see  quantitative differences. The continuum value 
of $\rho_{IR}/\sigma^{3/2}$, obtained with a quadratic fit, is 0.50(1) for
TI action and 0.71(2) for Wilson action, i.e.  they differ by factor 1.4.
We should make a comment on the different procedures of calculation
of $\rho_{IR}/\sigma^{3/2}$ for Wilson action 
in the present paper and in Refs.~\cite{Bornyakov:2001ux,ITEP}.
One difference is that in these papers 
only subsets of the full ensembles of gauge field configurations used in 
the present paper were employed. Another, more important difference is
that, in these earlier papers values of $\sigma$ from literature 
were used, while here we are using values of $\sigma$ calculated, as described 
in the previous section, i.e. on the same set of configurations on which the 
monopole density was calculated. 

The observed difference in  $\rho_{IR}$ measured for TI and Wilson actions,
though not being drastically large, means that the present definition of IR 
density is not universal. This makes it difficult to ascribe to it a meaning 
as a physical, gauge invariant density. It is evident that the source of
the discrepancy in the values of $\rho_{IR}$ might be the 
appending of UV monopoles, i.e. small loops, to IR monopole clusters. Since 
the TI action suppresses UV degrees of freedom stronger than 
Wilson action, it is natural to expect that this additional length assigned to
the IR cluster is smaller for TI action. Whether this is the only reason 
deserves further investigation.

The density of UV monopoles is reduced much more substantially, roughly by 
a factor 2.5, as can be seen from Fig.~\ref{fig:mondens2} (right). We can say  
that the TI action indeed suppresses (part of) the UV degrees of freedom.
\begin{figure}[hptb]
\begin{center}
\leavevmode
\hbox{\hspace{-1cm}
\epsfxsize=8.5truecm \epsfysize=6.8truecm \epsfbox{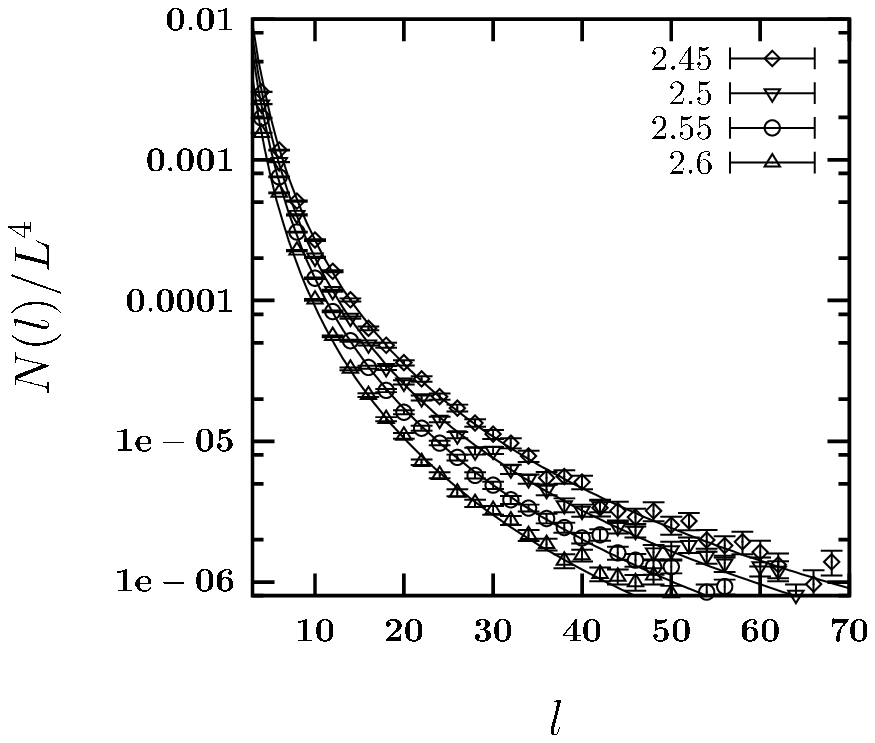}
\epsfxsize=8.5truecm \epsfysize=6.8truecm \epsfbox{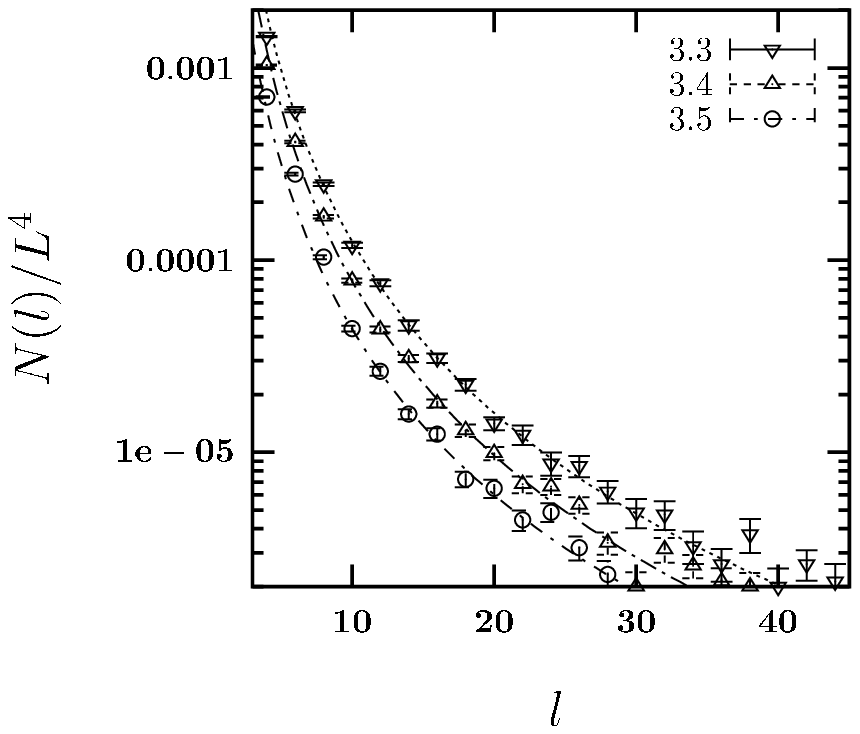}     }
\end{center}
\vspace{-1cm}
 \caption{The small (UV) clusters' length distribution in MAG for 
 the Wilson action at various $\beta$ (left)
 and for the TI action at various $\beta_{imp}$ (right). Curves are fits to 
Eq. (\ref{spectrum})}
\label{fig:mondist1}
\end{figure}
As generally expected for an improved action, one can also see earlier and 
faster convergence to the continuum limit.

As it has been mentioned above, the UV monopole density diverges.
It is natural to ask to which power of $a^{-1}$ this divergence is compatible.
It was first found in \cite{Bornyakov:2001ux}  
that for the Wilson action $\rho_{UV} \sim 1/a$. 
In Ref. \cite{ITEP} this was confirmed with higher confidence.
Fig.~\ref{fig:mondens2} (right) shows $a \rho_{UV}/\sigma$ for both actions. 
For TI action this ratio seems to rapidly converge to a finite value in the 
continuum limit as soon as $\sqrt{\sigma~a^2} < 0.25$ . This implies that 
$\rho_{UV} \sim 1/a$ also for TI action. Note, that the convergence for the 
TI action is faster than for the Wilson action, as can be seen from 
Fig.~\ref{fig:mondens2} (right).
If the existence of a reasonable continuum limit 
in the latter case would be confirmed, then only for $\sqrt{\sigma~a^2} << 0.1$,
and the limit value would be markedly larger than for the improved action.
In any case, the data clearly show that the UV monopole 
density and therefore the total density of monopoles is not universal.

The TI improved action has corrections of order $O(\alpha_s a^2)$ and $O(a^4)$
while the Wilson action has corrections of order $O(a^2)$. Thus it is natural to
expect that some contribution of lattice artifacts to the monopole density is 
suppressed in case of the improved action. Now we are able to conclude that a 
considerable part of the UV monopoles in the Wilson action case (more than 50 \%)
are lattice artifacts. In contrast to this, we notice that IR monopoles 
are not much affected by lattice artifacts~\footnote{The IR density is 
definitely affected by the roughness of monopole currents inside the largest 
cluster, and hence not completely free of discretization artefacts.}.
Whether the value of $\rho_{IR}$ obtained with the improved action is already 
the final universal one is an open question. This should be checked in simulations 
with other improved actions. 

The interesting, yet unanswered question is the physical role of UV monopoles. 
It was found in \cite{Hart:1997vb,Hart:1999gr} and then confirmed with higher 
precision in \cite{ITEP} that the number $N(l)$ of  small clusters 
of length $l$ falls like
\beq
\frac{N(l)}{L^4} = c(\beta)/l^\gamma \, ,
\label{spectrum}
\eeq
where $\gamma \approx 3$. 
The value $\gamma = 3$ was shown to be in agreement with percolation 
theory and also to be derivable within the polymer approach to the field theory 
of free or Coulomb-like interacting scalar particles \cite{Chernodub:2002tw}. 
Our data for both actions are also in agreement with relation (\ref{spectrum}),
with values of parameter $\gamma$ close to 3, as can be seen from 
Fig.~\ref{fig:mondist1} and Table~\ref{t4}. 
\begin{table}[hptb]
\begin{center}
\caption{Parameters of the fits to Eq. (\ref{spectrum}) of the small clusters's 
length distribution in MAG for both actions. Respective fit ranges 
($l_{min}, l_{max}$) are also shown.}\label{t4} 
\vspace{.3cm}
\setlength{\tabcolsep}{0.55pc}
\begin{tabular}{ccccc|ccccc}
$\beta$&$\gamma$& $c(\beta)$&$l_{min}$&$l_{max}$&$\beta_{imp}$&$\gamma$& $c(\beta)$&$l_{min}$&$l_{max}$  \\ 
\hline
2.45& 2.93(2) &0.23(2) &10&70&3.3&2.97(4)&0.12(1)&8 &40 \\
2.50& 2.99(2) &0.20(1) &12&50&3.4&3.01(7)&0.08(2)&10&30    \\
2.55& 3.08(2) &0.17(1) &10&50&3.5&2.87(9)&0.03(1)&10&30  \\
2.60& 3.11(6) &0.12(2) &14&40&&&&&    \\
\hline
\end{tabular}
\end{center}
\end{table}
%


Let us come back to observation that  $\rho_{UV} \sim \frac{1}{a}$.
This implies that magnetic currents from small clusters 
have a finite density per unit of 2D volume rather than per unit of 3D 
volume which is actually the case for the IR magnetic currents.
It has been recently verified that the density of $P$--vortices
in the indirect $Z(2)$ center gauge is finite in the continuum 
limit \cite{Gubarev:2002ek}.
On the other hand, it is known that in this gauge $P$--vortices and monopoles
are highly correlated \cite{DelDebbio:1997ke,Kovalenko:2004iu}. 
It is then natural
to assume that magnetic currents belonging to the small clusters ``populate'' 
the surfaces formed by $P$--vortices with some constant density. 
The strong reduction of the density of UV monopoles in the case of TI 
action in comparison with Wilson action suggests that the density of 
$P$--vortices will be suppressed, too. This should be checked in a future 
calculation. 

\begin{figure}[htbp]
\begin{center}
\hspace{-1.2cm}\includegraphics[width=8.4cm,angle=0]{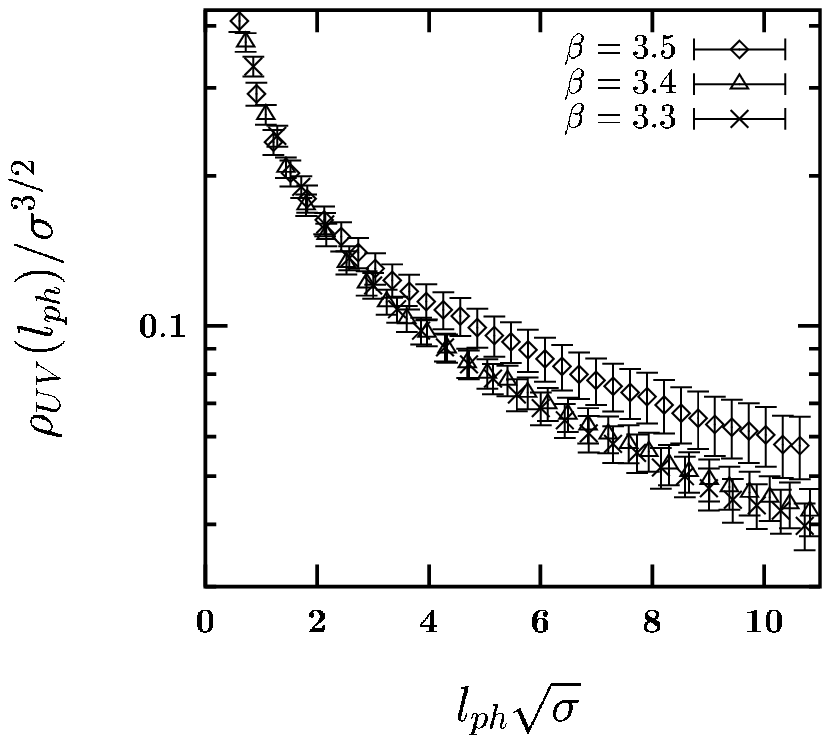}
\includegraphics[width=8.4cm,angle=0]{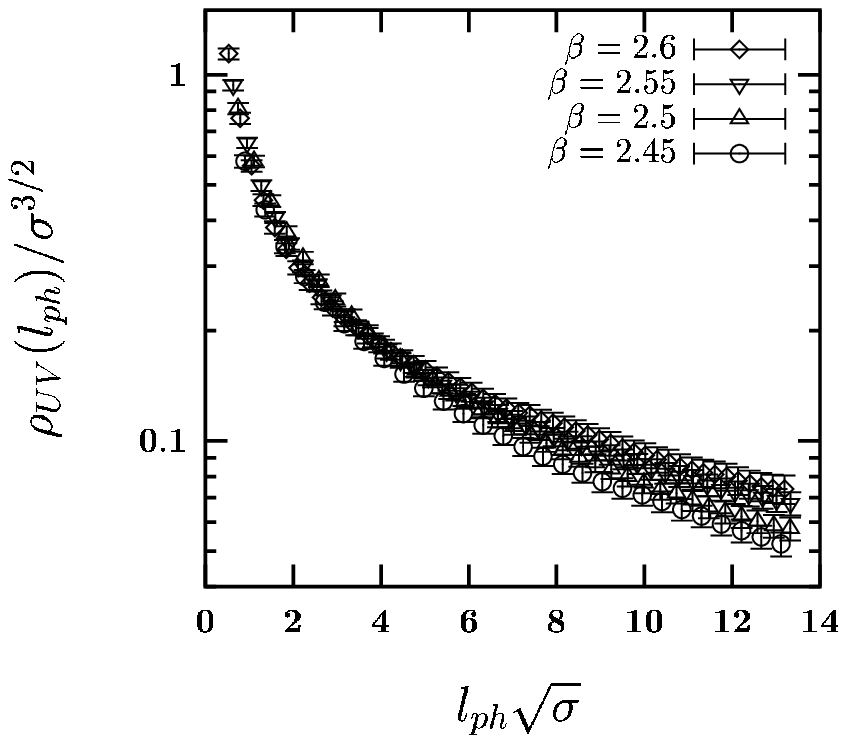}
\end{center}
\vspace{-1cm}
 \caption{The ``constrained'' UV monopole density derived from the small 
 clusters' length distribution Eq. (\ref{newdens}) for TI (left) and 
Wilson (right) actions.}
\label{fig:newdens1}
\end{figure}

\begin{figure}[htbp]
\begin{center}
\hspace{-1.2cm}\includegraphics[width=8.4cm,angle=0]{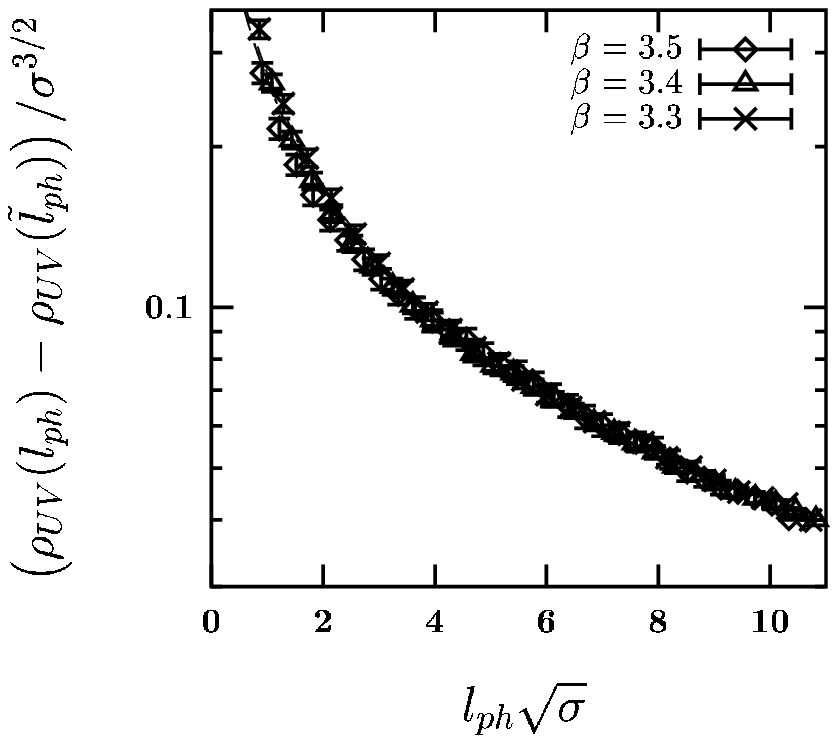}
\includegraphics[width=8.4cm,angle=0]{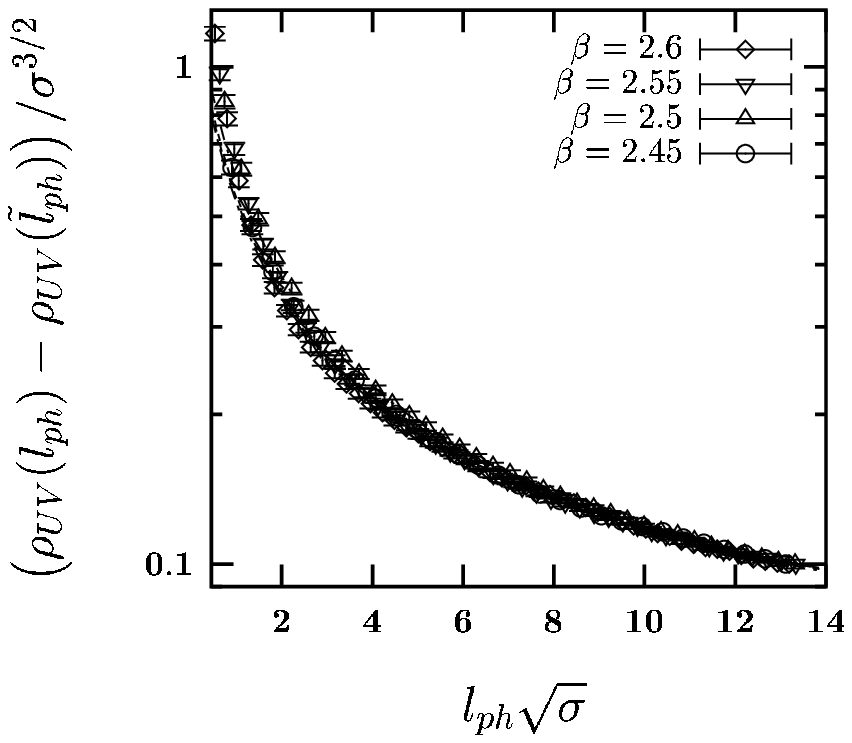}
\end{center}
\vspace{-1cm}
 \caption{The ``constrained'' UV monopole density derived from the small 
 clusters' length distribution and corrected for ambiguous clusters according
 to Eq. (\ref{diffnewdens}), for TI (left) and Wilson (right) actions. 
 The curves show fits by Eq. (\ref{fitnewdens}). 
 The data were lifted by a constant for better readability of the figures.}
\label{fig:newdens2}
\end{figure}

We now introduce a regularized UV monopole density by
summation of the average number of small clusters of length $l$
per lattice volume, {\it i.e.} $N(l)$, multiplied by the length $l$.
In this definition we shall exclude clusters below a certain (minimal) 
length scale $l_{ph}$ specified in natural units:
\begin{equation}
\rho_{UV}(l_{ph}) = \frac{1}{4L^4 a^3}\sum_{l \ge l_{ph}/a}  N(l)~l\,,
\label{newdens}
\end{equation}
We call this UV monopole density ``constrained density''. This definition
counts all monopole currents in small clusters with a length above or equal 
$l_{ph}$, and the emerging density depends on it as a parameter.
Thus, very small loops sensitive to the ultraviolate cutoff are excluded.
In Fig.~\ref{fig:newdens1} we show $\rho_{UV}(l_{ph})/\sigma^{3/2}$ as
a function of $l_{ph}$ for both actions. One can see that scaling is 
very good as long as $l_{ph}\sqrt{\sigma} < 2$ in the case of the TI 
action and $l_{ph}\sqrt{\sigma} < 4$ in the case of Wilson action.
In general, scaling becomes worse at larger values of $l_{ph}\sqrt{\sigma}$. 
This might be the consequence of the IR cluster splitting discussed above
which underlies the splitting into UV and IR monopoles: some large clusters 
which actually should belong (are akin) to IR monopole clusters were counted 
as small ones because of trivial winding. 
Although such clusters are relatively seldom their number increases
with increasing $\beta$. To exclude the effect of these ambiguously 
identified ``UV'' clusters we subtract the contribution of clusters with
$l > \tilde{l}_{ph} = c/\sqrt{\sigma} $ and plot in  
Fig.~\ref{fig:newdens2} the difference
\begin{equation}
\rho_{UV}(l_{ph}) - \rho_{UV}(\tilde{l}_{ph}) 
\label{diffnewdens}
\end{equation}
with coefficients $c \approx 13.6$ and $c \approx 10.65$
for Wilson action and TI action, respectively. 
One can see that now scaling is uniformly well satified for all lower 
cutoffs $l_{ph}< \tilde{l}_{ph}$ except very small ones.
Thus we come to the conclusion that the UV monopole density derived from the
small cluster density $N(l)$ as defined in eq. (\ref{newdens}), {\it i.e.} 
when clusters close to the cutoff scale are excluded,
shows good scaling, {\it i.e.} is independent of $a$ similar to 
the IR monopoles density (derived from the IR clusters). 

Using eq. (\ref{spectrum}) with $\gamma=3$ we get
\begin{equation}
\rho_{UV}(l_{ph}) =\frac{c(\beta)}{4 a^3} \psi^\prime (l_{ph}/a), \,\,\,\, 
\psi(z) = \frac{\Gamma^\prime(z)}{\Gamma(z)}
\end{equation}
We fitted the data in Fig.~\ref{fig:newdens2} by the function 
\begin{equation}
c_1 \psi^\prime (l_{ph}/a) + c_2
\label{fitnewdens}
\end{equation}
and found that this function describes well our data, especially for the 
TI action case, with constants $c_i$ only weakly dependent on $\beta$. 
This implies that 
for small $a$ (large $l_{ph}/a$) the density behaves
as $\rho_{UV}(l_{ph}) \sim \sigma/l_{ph}$ since $\psi^\prime(z) \sim 1/z$ for 
large
$|z|$. This fact is in agreement with $1/a$ behavior of $\rho_{UV}$ 
we argued for, since it can be unambiguously regulated by the assignment 
$\rho_{UV}(l_{ph}=4a) \rightarrow \rho_{UV}$.
\subsection{The Laplacian Abelian gauge (LAG)}
\label{sect:monopole_density_LAG}
To calculate the lowest eigenvector of the covariant adjoint Laplacian operator 
(\ref{eq:Laplacian})
we used the Arnoldi algorithm \cite{Arnoldi}. This algorithm, as well as others used
to solve this problem, requires large memory increasing fast with lattice size.
Therefore, the measurements in LAG for Wilson action have been made 
on smaller lattices than shown in Table~\ref{t2} for $\beta=2.45, 2.55, 2.6$, 
but with large enough physical size,
which was never smaller than 1.4 fm. In case of the TI action we made 
measurements in LAG only for four values of the coupling constant, 
because this proved to be enough for our purposes of comparison with MAG.

\begin{figure}[hpbt]
\hspace{-1.5cm}
\includegraphics[width=9.0cm,height=7cm,angle=0]{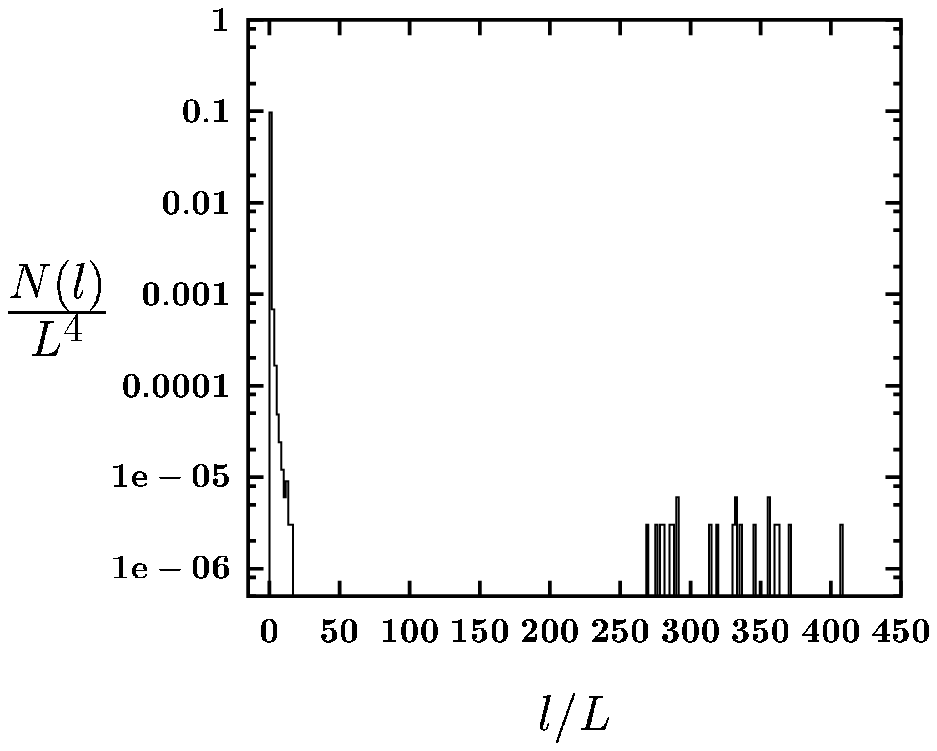}
\hspace{-1.cm}
\includegraphics[width=9.0cm,height=7cm,angle=0]{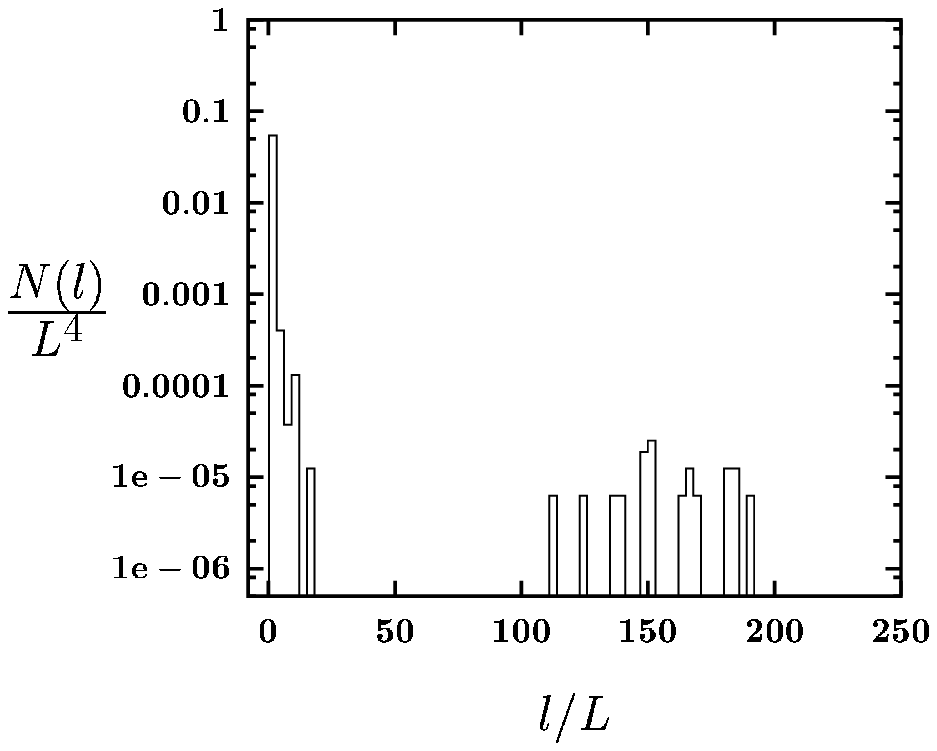}
\caption{The monopole cluster length distribution $N(l)$ in LAG
at $\beta=2.6$ for the Wilson action (left) and at $\beta_{imp}=3.5$ for
the TI action (right). }
\label{fig:hist-lag}
\end{figure}

\begin{figure}[hpbt]
\begin{center}
\hspace{-1.0cm}\includegraphics[width=8.4cm,angle=0]{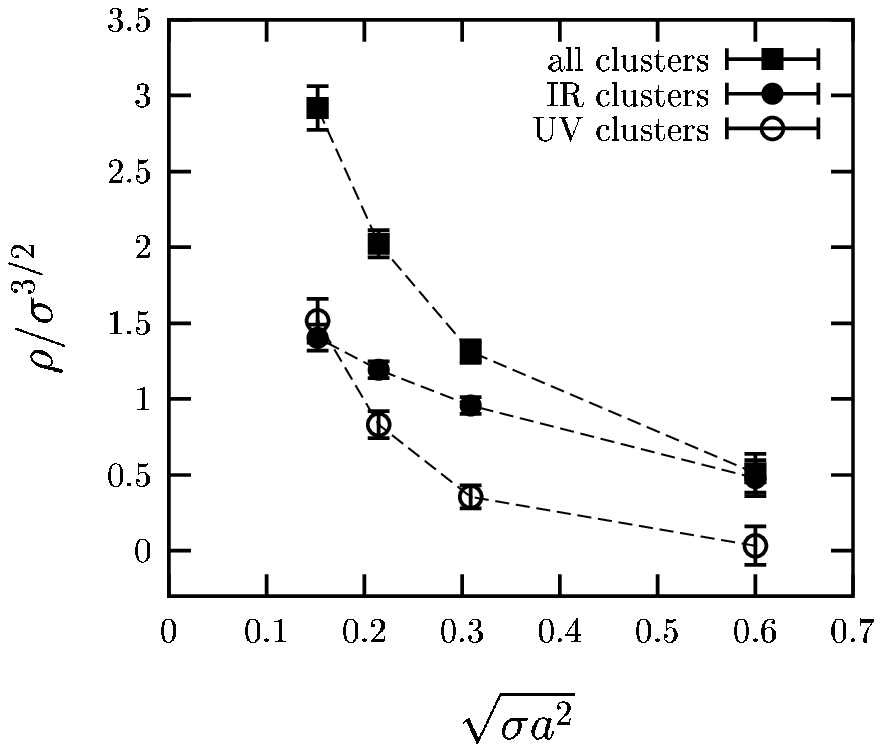}
\hspace{-0.5cm}\includegraphics[width=8.4cm,angle=0]{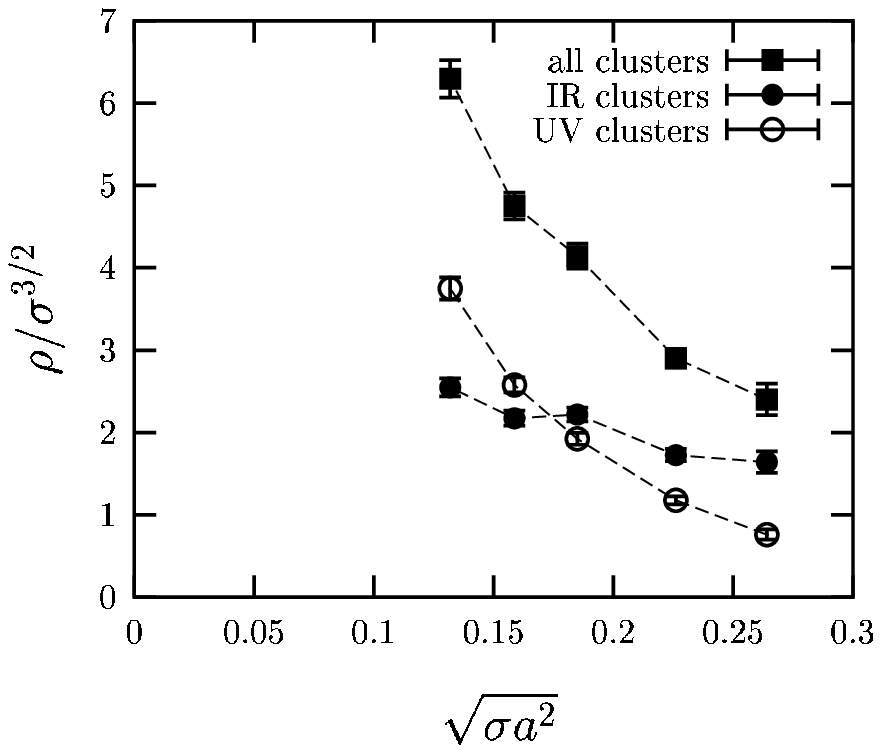}
\caption{Various monopole densities in LAG for TI action (left) and for Wilson
action (right). Lines are drawn to guide the eye.}
\label{fig:lag-dens}
\end{center}
\end{figure}

We first present the cluster length distribution for the two actions
in Fig.~\ref{fig:hist-lag}. One can see that in LAG the separation
into IR and UV clusters works very well. We note that there are only 
rare cases of clusters with nontrivial winding even for most fine lattices.
Our results for various densities are presented in Fig.~\ref{fig:lag-dens}.
We found that the total monopole density in LAG is substantially higher than 
in MAG in agreement with earlier observations \cite{vanderSijs:1997hi} made for 
Wilson action.
We further looked at IR and UV densities separately and found that increasing 
of the density is true for both of them.  For our
finest lattice these densities in LAG are 2--3 times higher than in MAG.
It is not clear from our data whether the IR density in LAG converges to 
a finite value as it is the case in MAG.  This is most probably due to a 
substantial (more fractal) admixture of lattice artefacts to the IR clusters.
We also observed that 
the UV monopole density $\rho_{UV}$ is much more strongly diverging
in LAG than in MAG and is compatible with $\rho_{UV} \sim 1/a^2$ rather than 
with $\rho_{UV} \sim 1/a$, as observed in MAG. Accordingly, the constrained 
density $\rho_{UV}(l_ph)$ was found to be divergent as 
$\frac{\sqrt{\sigma}}{a l_{ph}}$. The small clusters' length distribution 
was fitted to eq.(\ref{spectrum}) with a resulting parameter $\gamma$ in the 
range $2.7 - 2.8$, i.e. significantly smaller than for MAG.  

\section{The Abelian string tension in MAG}  
\label{abstr}

\begin{figure}[htbp]
\begin{center}
\leavevmode
\hbox{
\epsfxsize=10.0truecm \epsfysize=9.0truecm \epsfbox{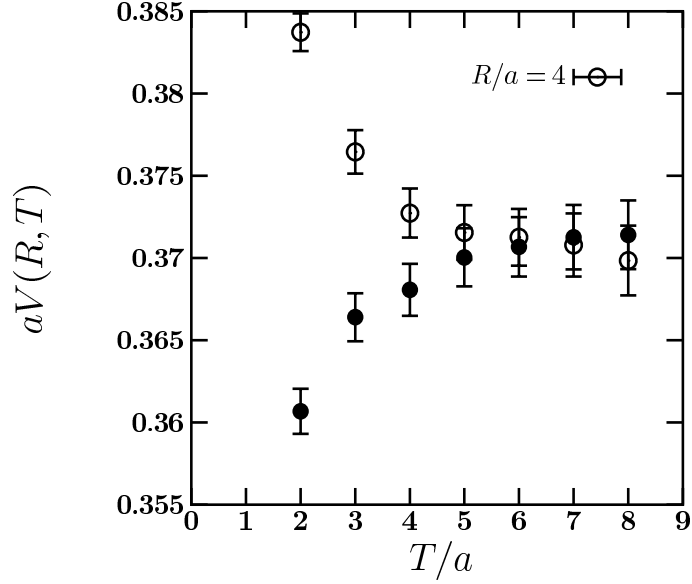}}
\end{center}
\vspace{-1cm}
 \caption{The Abelian potential 
 $V_{ab}(R,T)$ in MAG 
 at $\beta_{imp}=3.3$ vs. $T/a$ for $R/a=4$ for 
 3 (open circles) and 100 (full circles) smearing sweeps.} 
\label{fig:ab-pot-b33}
\end{figure}

To estimate the Abelian string tension we calculate the Abelian potential
$V_{ab}(R)$ using (spatially) smeared Abelian Wilson loops $W_{ab}(R,T)$. As 
usually $V_{ab}(R)$ is defined as a limit
\begin{equation}
V_{ab}(R) = \lim_{T \to \infty} V_{ab}(R,T),
\end{equation}
where the potential estimator $V_{ab}(R,T)$ is
\begin{equation}
aV_{ab}(R,T) = -\mbox{log}\frac{W_{ab}(R,T+a)}{W_{ab}(R,T)}\,. 
\end{equation}
It is important to check that $V_{ab}(R)$ is unique, i.e. independent of
the operators used to create the Abelian flux tube. One can get different such
 operators varying  the number of smearing sweeps $N_{sm}$. 
In Fig.~\ref{fig:ab-pot-b33}
we show $V_{ab}(R,T)$ for $N_{sm}=3$ and 100. For large number of sweeps 
the behavior of $V_{ab}(R,T)$ clearly shows absence of the positivity.
For small number of sweeps the behavior is similar to the case when positivity
is fulfilled.
This is presumably due to higher excitations: in this case they 
are not suppressed for small $T/a$. Lack of positivity for gauge dependent 
correlators in covariant gauges was discussed recently in \cite{Ogilvie}. 
The most important for us observation to be read off from 
Fig.~\ref{fig:ab-pot-b33} is that at large enough $T$ results
agree with each other. This implies that $V_{ab}(R)$ is indeed defined uniquely.

\begin{figure}[htbp]
\begin{center}
\leavevmode
\hbox{\hspace{-1cm}
\epsfxsize=8.0truecm \epsfysize=7.4truecm \epsfbox{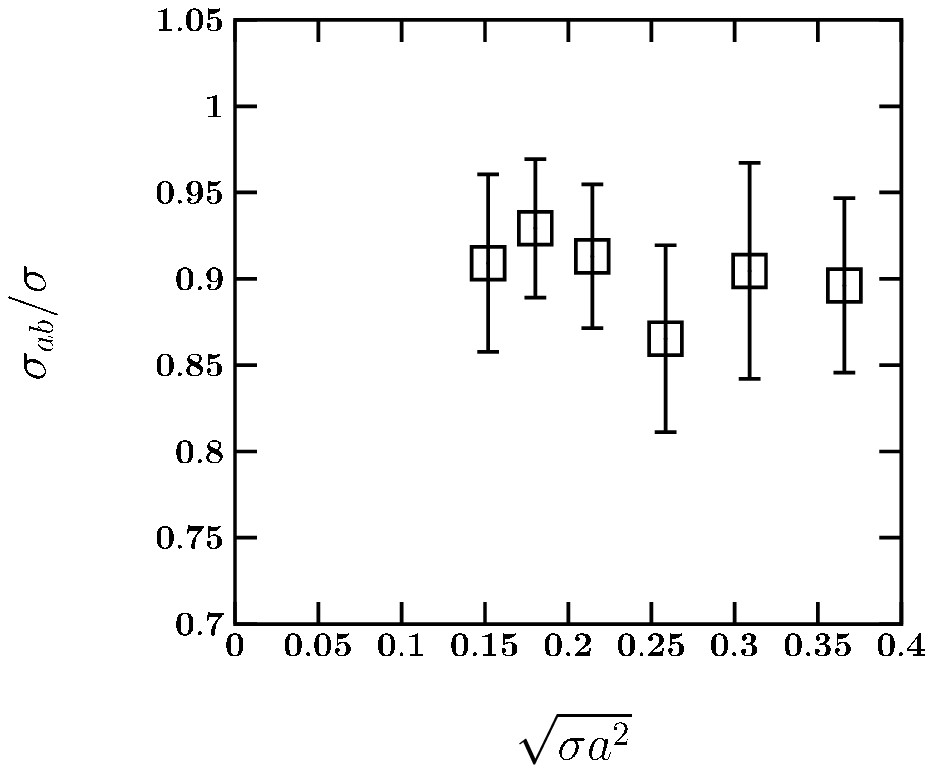}
\vspace{-1cm}\epsfxsize=8.0truecm \epsfysize=7.4truecm \epsfbox{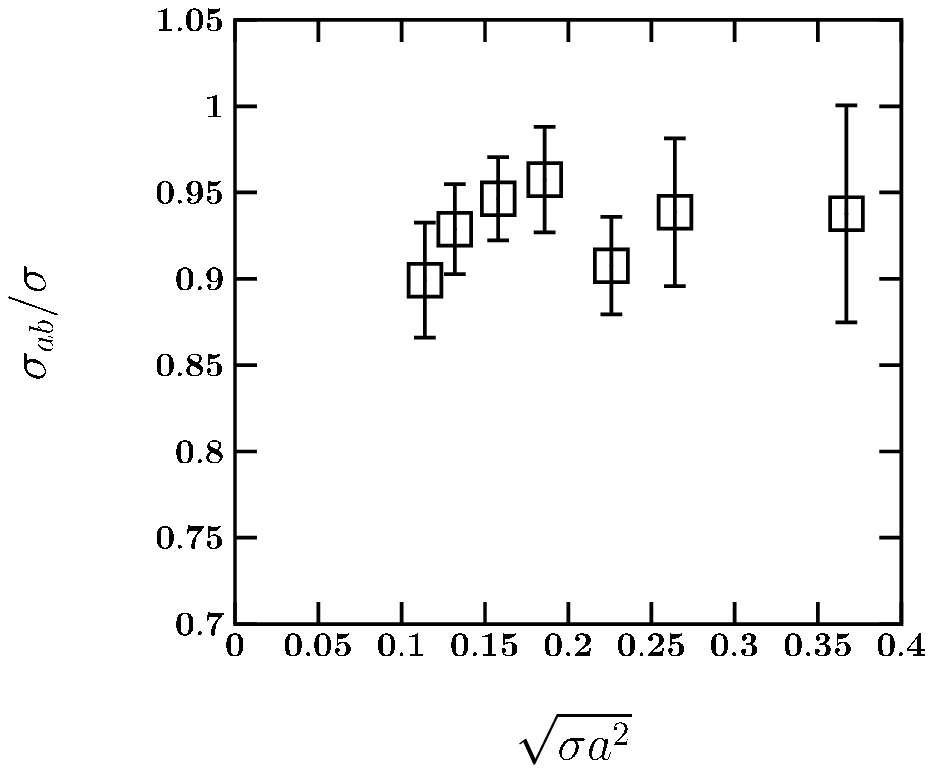}} 
\end{center}
\vspace{-1cm}
\caption{Ratio between the Abelian string tension (in MAG) 
and the non-Abelian string tension vs. lattice spacing for 
TI action (left)  and for Wilson action (right).}
\label{fig:ab-str-impr}
\end{figure}

We found that the behaviour of the ratio $\sigma_{ab}/\sigma$ for TI action
is qualitatively similar to that for the Wilson action, see 
Fig.~\ref{fig:ab-str-impr}.
For both actions the ratio is between 0.9 and 0.95 for all considered values 
of lattice spacing. We thus 
definitely confirm the universality of Abelian
dominance in the continuum limit. Our results, due to large statistical errors,
coming mainly from the determination of the non-Abelian string tension, do not
allow to determine precisely the continuum limit of the ratio 
$\sigma_{ab}/\sigma$. 

\section{Summary}
\label{summ}
In this paper two important questions on the properties of the
Abelian projection were addressed: universality and gauge dependence.
Comparing results obtained with Wilson and TI actions on lattices
with varying lattice spacing we confirmed that the Abelian dominance
passes the universality check in MAG. Moreover, this universality holds in the
continuum limit. We found that in the continuum limit the ratio
$\sigma_{ab}/\sigma$ seems to be in the range from 0.90 to 0.95. No 
convergence to 1 was observed contrary to our earlier results 
\cite{Bornyakov:2001ux} seen with smaller statistics.
We have not yet accomplished the check of the monopole dominance
universality for the string tension although our preliminary results confirm it. 
They will be published elsewhere. 

For the monopole density we found qualitative similarities:
for both actions $\rho_{IR}$ is finite in the continuum limit, while
$\rho_{UV}$ is divergent as $1/a$. On the quantitative level we found
a violation of universality for the densities. This implies that an UV contribution
might be substantial in
the measured $\rho_{IR}$, or in other words, $\rho_{IR}$ needs to be properly 
renormalized. 

We further introduced a constrained UV density $\rho_{UV}(l_{ph})$, 
eq. (\ref{newdens}) which is determined by counting only monopole loops 
longer than some physical length $l_{ph}$. We found that this density
scales properly in the continuum limit:
\begin{equation}
\rho_{UV}(l_{ph}) = \frac{c\sigma}{l_{ph}}\,,
\end{equation}
where the coefficient $c$ is independent of $a$ but 
has a different value for Wilson and TI actions, {\it i.e.} is non-universal.

A comparison of the monopole densities in MAG and LAG, made for both actions,
revealed that both $\rho_{IR}$ and $\rho_{UV}$ are 2-3 times higher
in LAG. It is not clear from our data whether $\rho_{IR}$ in LAG is finite 
in the continuum limit. The UV component $\rho_{UV}$ behaves like $\sim 1/a^2$ 
in LAG contrary to the divergence $\sim 1/a$ found in MAG.  
This adds to other doubts expressed in the literature~\cite{Langfeld} 
about the usefulness of the LAG.
Thus, the maximally Abelian gauge turns out to be particularly suited to the 
separation of IR and UV degrees of freedom. 

{\bf Acknowledgements}
This work was partially supported by grants RFBR 05-02-16306, RFBR 04-02-16079,
RFBR-DFG 03-02-04016, DFG-RFBR 436 RUS 113/739/0, and by the EU Integrated 
Infrastructure Initiative Hadron Physics (I3HP). V.B. is grateful to
the colleagues at ITP, Kanazawa University  and at NIC, DESY-Zeuthen,
where his contribution to this paper was worked out, for kind hospitality.
When this work has been started, E.-M. I. was supported by the
Ministry of Education, Culture and Science of Japan and
was enjoying the hospitality of H. Toki and the Theory group at RCNP,
Osaka University. E.-M. I. and M. M.-P. acknowledge the present support by
the DFG through the Forschergruppe FOR 465.

\end{document}